# Software Effort Estimation using Radial Basis and Generalized Regression Neural Networks

Prasad Reddy P.V.G.D, Sudha K.R, Rama Sree P and Ramesh S.N.S.V.S.C

**Abstract -**Software development effort estimation is one of the most major activities in software project management. A number of models have been proposed to construct a relationship between software size and effort; however we still have problems for effort estimation. This is because project data, available in the initial stages of project is often incomplete, inconsistent, uncertain and unclear. The need for accurate effort estimation in software industry is still a challenge. Artificial Neural Network models are more suitable in such situations. The present paper is concerned with developing software effort estimation models based on artificial neural networks. The models are designed to improve the performance of the network that suits to the COCOMO Model. Artificial Neural Network models are created using Radial Basis and Generalized Regression. A case study based on the COCOMO81 database compares the proposed neural network models with the Intermediate COCOMO. The results were analyzed using five different criterions MMRE, MARE, VARE, Mean BRE and Prediction. It is observed that the Radial Basis Neural Network provided better results.

**Index Terms—** Cognitive Simulation, Cost Estimation, Knowledge Acquisition, Neural Nets

—————————— ◆ ——————————

## 1 INTRODUCTION

IN algorithmic cost estimation [1], costs and efforts are predicted using mathematical formulae. The formulae are derived based on some historical data [2]. The best known algorithmic cost model called COCOMO (COnstructive COst MOdel) was published by Barry Boehm in 1981[3]. It was developed from the analysis of sixty three (63) software projects. Boehm proposed three levels of the model called Basic COCOMO, Intermediate COCOMO and Detailed COCOMO [3,5]. In the present paper we mainly focus on the Intermediate COCOMO.

### 1.1 Intermediate COCOMO

The Basic COCOMO model [3] is based on the relationship: Development Effort, $DE = a*(SIZE)^b$ where, SIZE is measured in thousand delivered source instructions. The constants a, b are dependent upon the 'mode' of development of projects. DE is measured in man-months. Boehm proposed 3 modes of projects [3]:

1. **Organic mode** – simple projects that engage small teams working in known and stable environments.

- *Prasad Reddy P.V.G.D,Dept of CSSE,AndhraUniversity,Vizag,AP*
- *Sudha K.R, Dept of EE,Andhra University,Vizag,AP,INDIA*
- *Rama Sree P,Dept of CSE,Aditya Engg. College,JNTUK,AP,INDIA*
- *Ramesh S.N.S.V.S.C,Dept of CSE,SSAIST,JNTUK,AP,INDIA*

2. **Semi-detached mode** – projects that engage teams with a mixture of experience. It is in between organic and embedded modes.
3. **Embedded mode** – complex projects that are developed under tight constraints with changing requirements.

The accuracy of Basic COCOMO is limited because it does not consider the factors like hardware, personnel, use of modern tools and other attributes that affect the project cost. Further, Boehm proposed the Intermediate COCOMO[3,4] that adds accuracy to the Basic COCOMO by multiplying 'Cost Drivers' into the equation with a new variable: EAF (Effort Adjustment Factor) shown in Table 1.

TABLE 1
DE FOR THE INTERMEDIATE COCOMO

| Development Mode | Intermediate Effort Equation |
|---|---|
| Organic | $DE = EAF * 3.2 * (SIZE)^{1.05}$ |
| Semi-detached | $DE = EAF * 3.0 * (SIZE)^{1.12}$ |
| Embedded | $DE = EAF * 2.8 * (SIZE)^{1.2}$ |

The EAF term is the product of 15 Cost Drivers [5] that are listed in Table 2. The multipliers of the cost drivers are Very Low, Low, Nominal, High, Very High and Extra High. For example, for a project, if RELY is Low, DATA is High, CPLX is extra high, TIME is Very High, STOR is High and rest parameters are Nominal then EAF = 0.75 * 1.08 *1.65*1.30*1.06 *1.0. If the category values of all the 15 cost drivers are "Nominal", then EAF is equal to 1.

TABLE 2
INTERMEDIATE COCOMO COST DRIVERS WITH MULTIPLIERS

| S. No | Cost Driver Symbol | Very low | Low | Nominal | High | Very high | Extra high |
|---|---|---|---|---|---|---|---|
| 1 | RELY | 0.75 | 0.88 | 1.00 | 1.15 | 1.40 | — |
| 2 | DATA | — | 0.94 | 1.00 | 1.08 | 1.16 | — |
| 3 | CPLX | 0.70 | 0.85 | 1.00 | 1.15 | 1.30 | 1.65 |
| 4 | TIME | — | — | 1.00 | 1.11 | 1.30 | 1.66 |
| 5 | STOR | — | — | 1.00 | 1.06 | 1.21 | 1.56 |
| 6 | VIRT | — | 0.87 | 1.00 | 1.15 | 1.30 | — |
| 7 | TURN | — | 0.87 | 1.00 | 1.07 | 1.15 | — |
| 8 | ACAP | — | 0.87 | 1.00 | 1.07 | 1.15 | — |
| 9 | AEXP | 1.29 | 1.13 | 1.00 | 0.91 | 0.82 | — |
| 10 | PCAP | 1.42 | 1.17 | 1.00 | 0.86 | 0.70 | — |
| 11 | VEXP | 1.21 | 1.10 | 1.00 | 0.90 | — | — |
| 12 | LEXP | 1.14 | 1.07 | 1.00 | 0.95 | — | — |
| 13 | MODP | 1.24 | 1.10 | 1.00 | 0.91 | 0.82 | — |
| 14 | TOOL | 1.24 | 1.10 | 1.00 | 0.91 | 0.83 | — |
| 15 | SCED | 1.23 | 1.08 | 1.00 | 1.04 | 1.10 | — |

The 15 cost drivers are broadly classified into 4 categories [3,5].
1. Product: RELY - Required software reliability
   DATA - Data base size
   CPLX - Product complexity
2. Platform: TIME - Execution time
   STOR—main storage constraint
   VIRT—virtual machine volatility
   TURN—computer turnaround time
3. Personnel: ACAP—analyst capability
   AEXP—applications experience
   PCAP—programmer capability
   VEXP—virtual machine experience
   LEXP—language experience
4. Project: MODP—modern programming
   TOOL—use of software tools
   SCED—required development schedule

Depending on the projects, multipliers of the cost drivers will vary and thereby the EAF may be greater than or less than 1, thus affecting the Effort [5].

## 2 PROPOSED NEURAL NETWORK MODELS

A neural network [14] is a massive parallel distributed processor made up of simple processing units, which has a natural propensity for storing experimental knowledge and making it available for use. It resembles the brain in two respects [4, 7, 11]:
1) Knowledge is acquired by the network from its environment through a learning process[15]
2) Interneuron connection strengths, known as synaptic weights, are used to store the acquired knowledge.

In this section we are going to present the two network models [12] used for the case study i.e. Radial Basis Neural Network(RBNN) and Generalized Regression Neural Network (GRNN).

### 2.1 Radial Basis Neural Network

Radial Basis Neural Network (RBNN) consists of two layers: a hidden radial basis layer of S1 neurons, and an output linear layer of S2 neurons [12]. A Radial Basis neuron model with R inputs is shown in Fig. 1. Radial Basis Neuron uses the radbas transfer function. The net input to the radbas transfer function is the vector distance between its weight vector w and the input vector p, multiplied by the bias b. (The || dist || box in this figure accepts the input vector p and the single row input weight matrix, and produces the dot product of the two.)The transfer function for a radial basis neuron is given in (1).

$$radbas(n) = e^{-n^2} \quad (1)$$

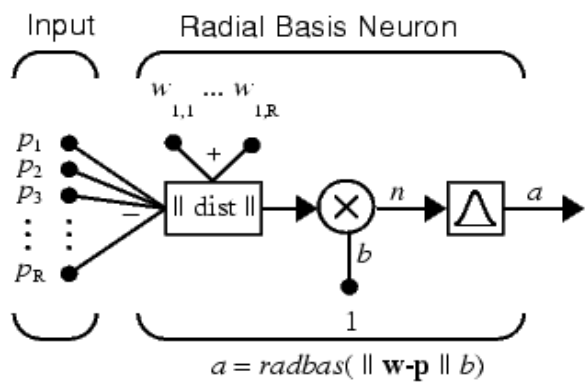

Fig. 1. Radial Basis neuron model

A plot of the radbas transfer function is shown in Fig.2.

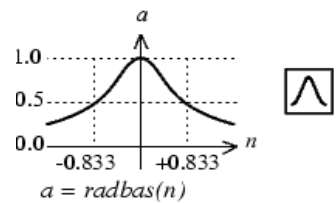

Fig. 2. radbas transfer function

The radial basis function has a maximum of 1 when its input is 0. As the distance between w and p decreases, the output increases. Thus, a radial basis neuron acts as a detector that produces 1 whenever the input p is identical to its weight vector w. The bias b allows the sensitivity of the radbas neuron to be adjusted. RBNN architecture is shown in Fig. 3.

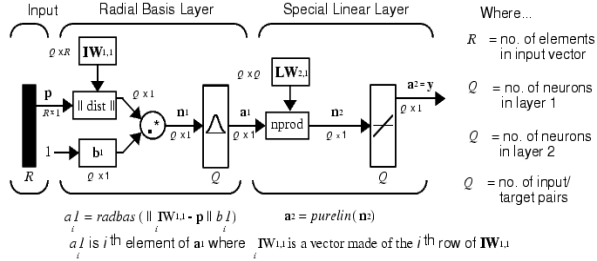

Fig. 3. Radial Basis Neural Network Architecture

The output of the first layer of this network net can be obtained using (2).

a{1} = radbas(netprod(dist(net.IW{1,1},p),net.b{1}))   (2)

If you present an input vector to this network, each neuron in the radial basis layer will output a value according to how close the input vector is to each neuron's weight vector. If a neuron has an output of 1, its output weights in the second layer pass their values to the linear neurons in the second layer. The second-layer weights LW $_{2,1}$ (or in code, LW{2,1}) and biases $b_2$ (or in code, b{2}) are found by simulating the first-layer outputs $a_1$ (A{1}), and then solving the linear expression (3).

$$[W\{2,1\}\ b\{2\}] * [A\{1\}; ones] = T \quad (3)$$

We know the inputs to the second layer (A{1}) and the target (T), and the layer is linear. We can use (4) to calculate the weights and biases of the second layer to minimize the sum-squared error.

$$Wb = T/[P; ones(1,Q)] \quad (4)$$

Here Wb contains both weights and biases, with the biases in the last column. There is another factor called SPREAD used in the network. The user chooses SPREAD, that is the distance an input vector must be from a neuron's weight. A larger SPREAD leads to a large area around the input vector where layer 1 neurons will respond with significant outputs. Therefore if SPREAD is small, the radial basis function is very steep, so that the neuron with the weight vector closest to the input will have a much larger output than other neurons. The network tends to respond with the target vector associated with the nearest design input vector.

### 2.2 Generalized Regression Neural Networks

A generalized regression neural network (GRNN) is often used for function approximation [8,9]. It has a radial basis layer and a special linear layer. The architecture for the GRNN is shown in Figure 4. It is similar to the radial basis network, but has a slightly different second layer.

Fig. 4. Generalized Regression Neural Network Architecture

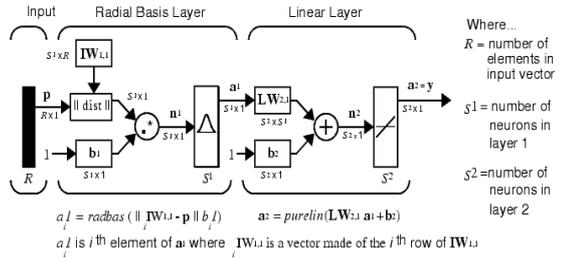

The first layer is just like that for Radial Basis networks. The second layer also has as many neurons as input/target vectors, but here LW{2,1} is set to T. Here the nprod box shown above produces $S_2$ elements in vector $n_2$. Each element is the dot product of a row of $LW_{2,1}$ and the input vector $a_1$. The user chooses SPREAD, the distance an input vector must be from a neuron's weight.

### 2.3 Advantages of Radial Basis Networks

1) Radial basis networks can be designed in a fraction of the time that it takes to train standard feed forward networks. They work best when many training vectors are available.
2) Radial Basis Networks are created with zero error on training vectors.

## 3 VARIOUS CRITERIONS FOR ASSESSMENT OF ESTIMATION MODELS

1. Mean Absolute Relative Error (MARE)

$$MARE\ (\%) = \frac{\sum f(R_E)}{\sum f} \times 100 \quad (5)$$

2. Variance Absolute Relative Error (VARE)

$$VARE(\%) = \frac{\sum f(R_E - meanR_E)}{\sum f} \times 100 \quad (6)$$

3. Prediction (n)

Prediction at level n is defined as the % of projects that have absolute relative error less than n.

4. Balance Relative Error (BRE)

$$BRE = \frac{|E - \hat{E}|}{\min(E, \hat{E})} \quad (7)$$

5. Mean Magnitude of Relative Error (MMRE)

$$MMRE(\%) = \frac{1}{N}\sum_{i=1}^{N} MRE_i \times 100 \quad (8)$$

Where $MRE = \left| \dfrac{\hat{E} - E}{\hat{E}} \right|$, N = No. of Projects

E = estimated effort   Ê = actual effort

Absolute Relative Error (RE) = $\left| \dfrac{\hat{E} - E}{\hat{E}} \right|$

A model which gives lower MARE (5) is better than that which gives higher MARE. A model which gives lower VARE is better than that which gives higher VARE [6]. A model which gives lower BRE (7) is better than that which gives higher BRE. A model which gives higher Pred (n) is better than that which gives lower Pred (n). A model which gives lower MMRE (8) is better than that which gives higher MMRE.

## 4  EXPERIMENTAL STUDY

The COCOMO81 database [5] consists of 63 projects data [3], out of which 28 are Embedded Mode Projects, 12 are Semi-Detached Mode Projects, and 23 are Organic Mode Projects. In carrying out our experiments, we have chosen the COCOMO81 dataset[13]. Out of 63 projects, randomly selected 53 projects are used as training data. A Radial Basis Network and Generalized Regression Network are created .The two networks are tested using the 63 dataset. For creating radial basis network, newrbe( ) is used and for creating generalized regression network, newgrnn( ) is used. We have used a SPREAD value of 0.94. The estimated efforts using Intermediate COCOMO, RBNN and GRNN are shown for some sample projects in Table 3.  The Effort is calculated in man-months. Table 4 and Fig.5., Fig.6., Fig.7., Fig.8., Fig.9., Fig.10. & Fig. 11. shows the comparisons of various models [10] basing on different criterions.

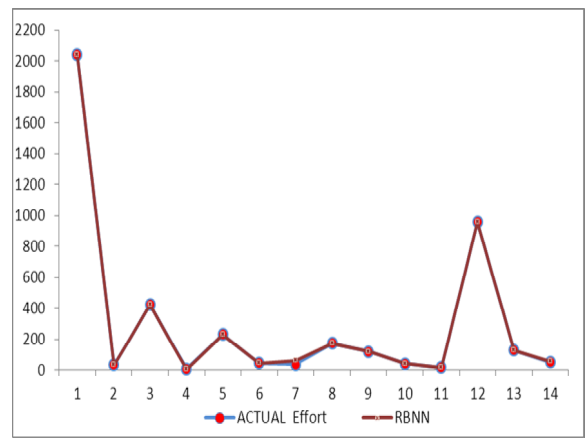

Fig. 5. Actual Effort  versus RBNN Effort

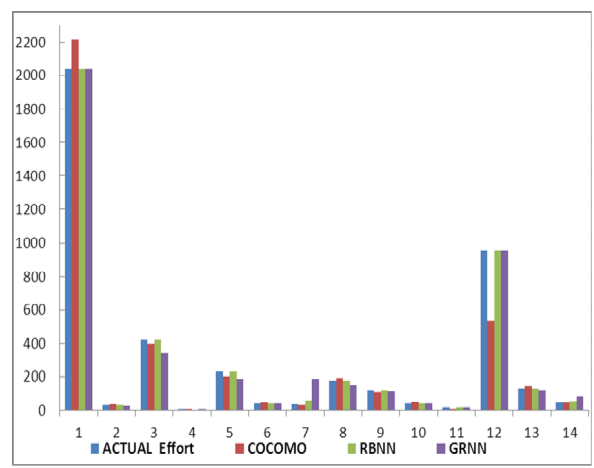

Fig. 6. Estimated Effort of various models versus Actual Effort

TABLE 3
ESTIMATED EFFORT IN MAN MONTHS OF VARIOUS MODELS

| Project ID | ACTUAL EFFORT | Estimated EFFORT using | | |
|---|---|---|---|---|
| | | COCOMO | RBNN | GRNN |
| 1 | 2040 | 2218 | 2040 | 2040 |
| 5 | 33 | 39 | 33 | 30 |
| 9 | 423 | 397 | 423 | 345 |
| 29 | 7.3 | 7 | 5.6 | 6.8 |
| 34 | 230 | 201 | 230 | 188 |
| 42 | 45 | 46 | 45 | 45 |
| 47 | 36 | 33 | 62 | 187 |
| 50 | 176 | 193 | 176 | 151 |
| 51 | 122 | 114 | 122 | 115 |
| 52 | 41 | 55 | 41 | 43 |
| 55 | 18 | 7.5 | 18 | 16 |
| 56 | 958 | 537 | 958 | 954 |
| 58 | 130 | 145 | 130 | 124 |
| 61 | 50 | 47 | 57 | 88 |

TABLE 4
COMPARISON OF VARIOUS MODELS

| Model | MARE (%) | VARE (%) | Mean BRE | MMRE (%) | Pred(40) (%) |
|---|---|---|---|---|---|
| Intermediate COCOMO | 19.45 | 4.97 | 0.22 | 18.60 | 87.3 |
| RBNN | 7.13 | 3.27 | 0.17 | 17.29 | 90.48 |
| GRNN | 14.19 | 4.44 | 0.35 | 34.61 | 84.13 |

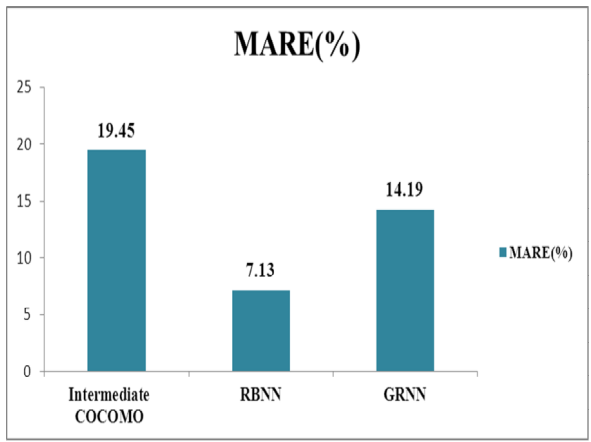

Fig. 7. Comparison of MARE against various models

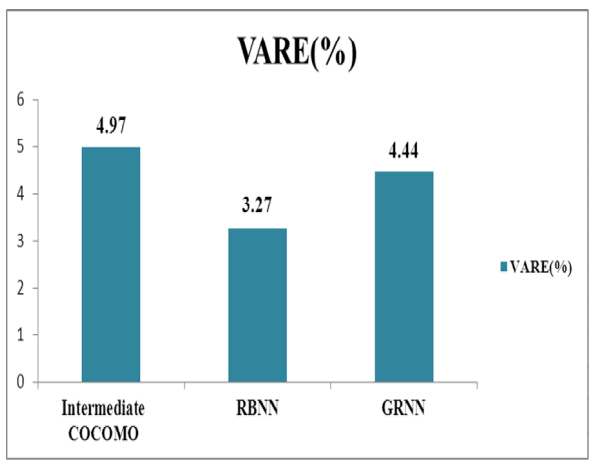

Fig. 8. Comparison of VARE against various models

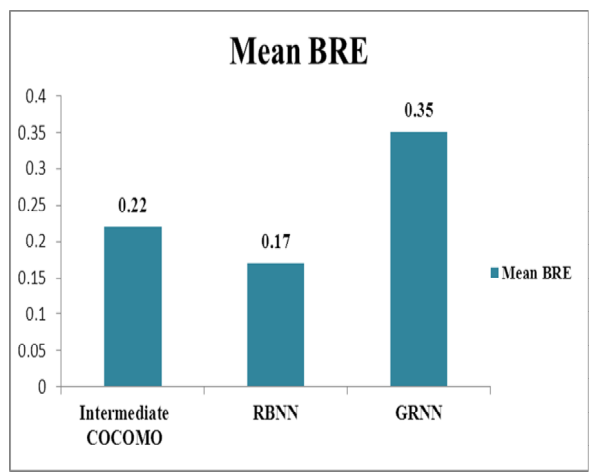

Fig. 9. Comparison of Mean BRE against various models

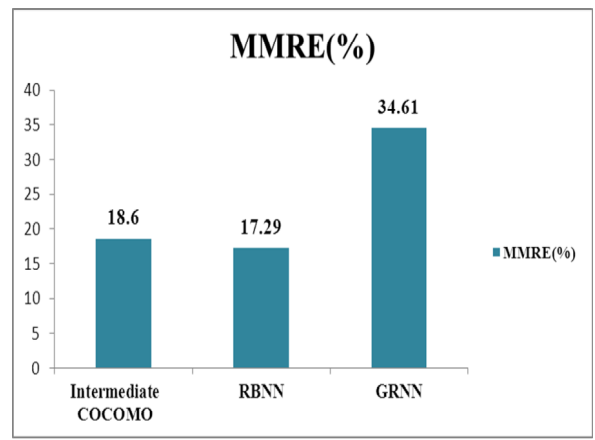

Fig. 10. Comparison of MMRE against various models

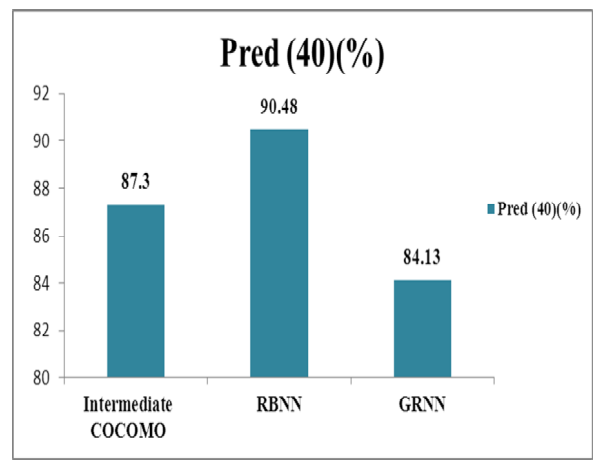

Fig. 11. Comparison of Pred(40) against various models

## 5 CONCLUSION

Referring to Table 4, we see that Radial Basis Neural Networks yields better results for maximum criterions when compared with the other models. Thus, basing on MARE, VARE, Mean BRE, MMRE & Pred(40) we come to a conclusion that RBNN is better than GRNN or Intermediate COCOMO. Therefore we proved that it's better to create a Radial Basis Neural Network for software effort prediction using some training data and use it for effort estimation for all the other projects.

## REFERENCES

[1] Ramil, J.F, Algorithmic cost estimation for software evolution, Software Engg. (2000) 701-703.
[2] Angelis L, Stamelos I, Morisio M, Building a software cost estimation model based on categorical data, Software Metrics Symposium, 2001- Seventh International Volume (2001) 4-15.
[3] B.W. Boehm, Software Engineering Economics, Prentice-Hall, Englewood Cli4s, NJ, 1981
[4] Dawson, C.W., "A neural network approach to software projects

effort estimation," Transaction: Information and Communication Technologies, Volume 16, pages 9,1996.

[5] Zhiwei Xu, Taghi M. Khoshgoftaar, Identification of fuzzy models of software cost estimation, Fuzzy Sets and Systems 145 (2004) 141–163

[6] Harish Mittal, Harish Mittal, Optimization Criteria for Effort Estimation using Fuzzy Technique, CLEI Electronic Journal, Vol 10, No 1, Paper 2, 2007

[7] Idri, A. Khoshgoftaar, T.M. Abran, A., "Can neural networks be easily interpreted in software cost estimation?," Proceedings of the IEEE International Conference on Fuzzy Systems, FUZZ-IEEE'02, Vol.: 2, 1162-1167, 2002

[8] Generalized Regression Neural Nets in Estimating the High-Tech Equipment Project Cost, 2010 Second International Conference on Computer Engineering and Applications, Bali Island, Indonesia March 19-March 21,ISBN: 978-0-7695-3982-9

[9] Mehmet Ali Yurdusev ; Mahmut Firat;Mustafa Erkan Turan "Generalized regression neural networks for municipal water consumption prediction" Published in: Journal of Statistical Computation and Simulation , Volume 80, Issue 4 April 2010 , pages 477 - 478

[10] Heiat, A., "Comparison of artificial neural network and regression models for estimating software development effort", Information and Software Technology 44 (15), 911–922, 2002.

[11] Dawson, C.W., "A neural network approach to software projects effort estimation," Transaction: Information and Communication Technologies, Volume 16, pages 9, 1996

[12] Validating and Understanding Software Cost Estimation Models based on Neural Networks, Ali Idri, Alain Abran, Samir Mbarki 0-7803-8482-2/04,2004 IEEE

[13] Donald J. Reifer, Barry W. Boehm, Sunita Chulani, "The Rosetta Stone: Making COCOMO 81 Files Work With COCOMO II", CROSSTALK, The Journal of Defense Software Engineering, Feb 1999

[14] Karunanithi, N., et al. "Using neural networks in reliability prediction", IEEE Software, pp. 53-59, 1992

[15] LiMin Fu, "Neural Networks in Computer Intelligence," Tata McGraw-Hill Edition 2003, pp.94-97.